# Effects of Ca substitution and the pseudogap on the magnetic properties of Y$_{1-x}$Ca$_x$Ba$_2$Cu$_3$O$_{7-\delta}$

S. H. Naqib[1,2*], J. R. Cooper[1], and J. W. Loram[1],

[1] *Department of Physics, University of Cambridge, J. J. Thomson Avenue, Cambridge CB3 OHE, UK*

[2] *Department of Physics, University of Rajshahi, Raj-6205, Bangladesh*

The effects of planar hole content, *p*, on the static magnetic susceptibility, $\chi(T)$, of Y$_{1-x}$Ca$_x$Ba$_2$Cu$_3$O$_{7-\delta}$ polycrystalline samples were investigated over a wide range of Ca (*x*) and oxygen contents. Non-magnetic Ca$^{2+}$, in the 3p$^6$ state, induces a Curie-like contribution to $\chi(T)$ that increases systematically and non-linearly with *x* but is almost independent of *p*. We argue that this arises from statistical clusters containing two or more nearest neighbor Ca atoms. We have again found that the pseudogap in the quasi-particle spectral weight appears abruptly below a planar hole content *p* = 0.190 ± 0.005.

# I. INTRODUCTION

The properties of high-$T_c$ copper oxide superconductors (HTS) in the normal and the superconducting (SC) states are highly dependent on the number of doped carriers per copper oxide plane, $p$, and one of the most widely studied phenomena is the so-called normal state pseudogap (PG) [1 - 4]. Effects of the pseudogap are observed in the $T$ - $p$ phase diagram of the cuprates over a certain doping range, extending from the underdoped (UD) to slightly overdoped (OD) regions. Many of the unusual properties can be interpreted in terms of a reduction in the quasi-particle density of states (DOS) near the chemical potential [1 - 6]. At present the experimental and the theoretical situations regarding the origin of the pseudogap are rather inconclusive [2, 4].

Here we report a systematic study of the static magnetic susceptibility, $\chi(T)$, of polycrystalline $Y_{1-x}Ca_xBa_2Cu_3O_{7-\delta}$ (Ca-Y123) over a wide range of $p$ as well as heat capacity results for a representative sample. One advantage of Ca ($x$) substitution is that the overdoped region can be studied, up to $p \sim 0.23$ with $x = 0.20$ [7, 8]. To our knowledge, detailed $\chi(T)$ measurements of Ca-substituted Y123 have not been reported so far. Based on experimental evidence [1, 3, 5] that $\chi(T)$ and $S/T$, where $S$ is the electronic entropy, show similar behavior, analysis of $\chi(T, p)$ data gives important information about the $T$- and $p$-dependences of the low-energy electronic density of states for this representative hole-doped cuprate. The variation of the DOS, $N(\varepsilon)$ with energy $\varepsilon$ is at the heart of any problem associated with the pseudogap, so $\chi(T, p)$ is a simple but powerful way of studying this quantity. The intrinsic spin part of $\chi(T, p)$, the Pauli spin susceptibility, $\chi_{spin}$ is a measure of the quasi-particle (QP) spectral density near the Fermi level. For Fermi liquids and in the absence of exchange enhancement, the Pauli spin susceptibility can be expressed as

$$\chi_{spin}(T) = \mu_B^2 <N(\varepsilon)>_T \qquad (1)$$

where $<N(\varepsilon)>_T \equiv \int N(\varepsilon)(\partial f/\partial \varepsilon)d\varepsilon$, is the thermal average of the DOS, $\mu_B$ is the Bohr magneton, and $f$ is the Fermi-function. Therefore, $\chi_{spin}$ at any particular temperature, $T$, represents the average of $N(\varepsilon)$ over an energy region $\sim \varepsilon_F \pm 2k_BT$ [5].

The main observations from the present study are: *(a)* we again find that the PG appears abruptly for compounds with $p < 0.19$. *(b)* non-magnetic Ca substitution induces a Curie term in $\chi(T)$, irrespective of the value of $p$. *(c)* we argue that this Ca induced Curie term is caused by clusters of two or more nearest neighbor $Ca^{2+}$ ions. If this is true, it has implications for establishing the number of mobile holes, $p$, in materials such as LSCO and Bi:2201, where the hole content is varied systematically by doping with a relatively large number of altervalent atoms.

## II. EXPERIMENTAL DETAILS

Polycrystalline samples of $Y_{1-x}Ca_xBa_2Cu_3O_{7-\delta}$ were synthesized by standard solid-state reaction methods from high-purity powders. Details of sample preparation and characterization can be found in refs. 9 - 11. A single sample was often used for each Ca content, the oxygen deficiency, $\delta$ and hence $p$ were varied by annealing at fixed temperatures and oxygen partial pressures and quenching the sample into liquid nitrogen [10]. Most normal and SC state properties including $E_g$ (the characteristic PG energy scale) of HTS are strongly dependent on $p$ which should therefore be determined as accurately as possible. We have used the room temperature thermopower, $S[290K]$ [12, 13] as well as the parabolic $T_c$ - $p$ [14] relation to determine $p$ for all samples. These two methods gave almost identical values of $p$. We measured $T_c$ using both resistivity and low-field AC susceptibility with an alternating field $H_{rms}$ = 0.1 Oe, and frequency $f$ = 333.3 Hz [13]. $T_c$ values obtained with these two methods agree to within 1 K for all the samples. Electron probe microanalysis (EPMA) was performed to verify the chemical composition and homogeneity of all samples.

*Quantum Design MPMS2* and *MPMS XL* SQUID magnetometers were used for the magnetic measurements reported here, for data from 5 - 400 K and 5 – 330 K respectively. A magnetic field of 5 T was applied, with field-linearity checks at 300 and 100 K, and the background signal from the sample holder was measured and subtracted from the raw data. Powder X-Ray diffraction patterns of all the samples studied here showed phase purity to within 1%. Raman spectroscopy studies suggested that the dominant magnetic impurity is $BaCuO_{2+z}$, because a peak at ~ 640 $cm^{-1}$ [15] with variable intensity was observed for different Ca-Y123 samples. Batch II of the 20%Ca-Y123 compound had a higher intensity peak than batch I and the compounds with lower Ca content.

Room-temperature electron spin resonance (ESR) at 9 GHz was used to search for unwanted magnetic impurity phases[†]. Two lines in the derivative spectra with peak to peak widths ~ 200 and ~ 1000 Gauss were usually visible, the ESR intensity generally being dominated by the broader line. Double integration of the ESR spectra and calibration with a known mass of a standard sample, pure $Y_2BaCuO_5$, gave values of the Curie constant ($C_{ESR}$) listed in Table 1. Although the ESR data showed that 0.3 - 1 % of the total number of Cu atoms were contained in impurity phases, *'a-priori'* there is some uncertainty [16, 17, 18] as to whether the broader line does indeed arise from $BaCuO_2$. Furthermore, because of the complex crystallographic and magnetic structure of $BaCuO_2$, [16, 17] it is not clear that all Cu spins would be detected by room temperature ESR. Later ESR experiments at lower T (and in fact two of the room temperature measurements shown in Table 1) were hampered by the partial decomposition of the powder samples giving a ferromagnetic component that was also seen in later SQUID magnetometer data but absent in the earlier SQUID data. Finally, partly as an extra check for magnetic impurity phases, the heat capacity, $C_v$, of a third highly overdoped 20%Ca-Y123 sample, with *S[290K] = - 6 μV/K* giving *p* = 0.218, was measured. This sample had a larger susceptibility than the two samples studied in detail e.g. at 300 K $\chi T$ = 0.143 emu-K/mole, but after correction for the amount of $BaCuO_2$ detected via the heat capacity, $\chi(T)$ is consistent with data for the other two 20% Ca samples.

Figs. 1 and 2 show plots of $\chi(T)$ and $\chi(T)T$ for the $Y_{1-x}Ca_xBa_2Cu_3O_{7-\delta}$ compounds at various values of *p* and readily illustrate two important points. The first one, related to the existence of the pseudogap (discussed in the next section), is that both $\chi(T)$ and $\chi(T)T$ become strongly *p*-dependent *only* for *p* < 0.19 (± 0.005). The second important finding is that there is a systematic growth of Curie-like behaviour in $\chi(T)$ with increasing Ca content. $\chi(T)$ for Ca-substituted Y123 shows features similar to Co- or Ni-substituted Y123 [6]. This is quite surprising because unlike Co and Ni, Ca is non-magnetic having a full outer shell, $3p^6$, in the doubly ionised state. Therefore no Curie-like contribution to the magnetization is expected. It has been suggested that oxygen vacancies are induced in the CuO$_2$ plane by increasing levels of Ca substitution [19]. Electron irradiaton studies [20] suggest that, if present, these vacancies would be strong perturbations similar to Zn/Cu substitution. We feel that oxygen defects are unlikely here because in contrast to the effect of strong in-plane scattering by Zn atoms, $T_{cmax}$ and the Hall angle [21, 22] are not altered very much by Ca substitution

---

[†] We are grateful to Prof. P. Monod, Laboratoire de Physique du Solide (UPR 5 CNRS), Paris, France, for performimg and interpreting the ESR measurements.

## III. DATA ANALYSIS

**Magnetic Susceptibility**

$\chi(T)$ of BaCuO$_{2+z}$ obeys a $C/(T-\theta)$ law between 120 and 300 K with $C$ and $\theta$ varying slightly with the oxygen content 2+z [23]. On comparing our annealing conditions with those in Ref. [23], we expect $C$ for the BaCuO$_{2+z}$ impurities in our Y123 samples to range from 0.384 emu-K/mole to 0.468 emu-K/mole and $\theta$ from 62 to 46 K as $p$ is changed from 0.22 to 0.11. Consistent with this, our data for all Y$_{1-x}$Ca$_x$Ba$_2$Cu$_3$O$_{7-\delta}$ samples with $p > 0.19$ and $T > 100$ K, can be fitted to:

$$\chi(T) = \chi_0 + C/T + C_B/(T-\theta) \qquad (2)$$

where $\chi_0$, $C$ and $C_B$ are constants and $\theta$ lies within the above range. Parameters obtained from such fits for the most heavily oxygenated samples and taking $\theta = 62$ K are given in Table 1. For the 0 and 5% Ca samples $C_B$ was zero within the experimental uncertainty of $\pm 5 \times 10^{-4}$ emu-K/mole. The values of $C_B$ for 20% Ca samples I and II correspond to 0.009 and 0.018 moles BaCuO$_{2+z}$ per mole Y$_{1-x}$Ca$_x$Ba$_2$Cu$_3$O$_{7-\delta}$ respectively, i.e. to 0.3 and 0.6 % of the total number of Cu atoms having spin $s \sim 1/2$ and are similar to the values of $C_{ESR}$. So it seems that for these samples the ESR signals did indeed arise from the BaCuO$_{2+z}$ impurities, and all the Cu spins in BaCuO$_{2+z}$ were detected. The variations in $C_B$ and $\theta$ with $z$ mentioned above have little effect on $\chi(T)$. For example in the worst case, the 20% Ca sample II, they only change $\chi(100)$ by 0.2 x 10$^{-4}$ emu-K/mole, or 4% of the much larger $C/T$ term. In summary, all our data for $p > 0.19$ are consistent with $C$ and $C_B$ being constant for a given Ca content within experimental uncertainty. We are therefore justified in subtracting these terms from the data for samples with $p < 0.19$ in order to explore the variation of the PG with $p$ in Ca-doped samples.

**Effect of the PG**

As mentioned in the Introduction, the striking correspondence between the spin-susceptibility and the electronic entropy, $S$, or more precisely $S/T$, is well documented [1, 5, 6]. This suggests that we may interpret our $\chi(T, p)$ data in terms of the DOS for the cuprates, in spite of the presence of strong electronic correlations in these compounds. The electronic entropy has a simple physical meaning, $S(T)/k_B$ counts the total number of thermally excited charge and spin excitations in the electronic spectrum in an energy window a few $k_BT$ wide, centered on the chemical potential. The quantity $\chi_{spin}T$ on the other hand, provides complementary information for the spin spectrum. Specifically, $k_B\chi_{spin}T = <\mu_z^2>$, where $<\mu_z^2>$ is the mean squared moment, and therefore $(k_B\chi_{spin}T)/\mu_B^2$ is a measure of the number of thermal spin excitations inside a similar energy window around the Fermi energy [5, 6]. For a V-shaped gap in the DOS, namely $N(\varepsilon) = N_0$ for $|\varepsilon - \varepsilon_F| > k_BE_g$ (where $E_g$ is the pseudogap energy scale expressed in degrees K) and $N(\varepsilon) = N_0|\varepsilon - \varepsilon_F| / k_BE_g$ for $|\varepsilon - \varepsilon_F| < k_BE_g$, $\chi_{spin}$ is given by Eqn. (1) with [5, 25]

$$<N(\varepsilon)>_T = N_0[1 - D^{-1}\ln\{\cosh(D)\}] \qquad (3)$$

where $D = E_g/2T$. Fig. 3 shows fits of $\chi(T, p)$ data [3] for pure Y123 samples to Eqns. (1) and (3), with a $p$-independent value of $N_0$, from 400 K to ~ $T_c + 30$ K (to avoid significant SC fluctuations near $T_c$). $E_g(p)$ values obtained from the fits are shown in Fig. 3 and also plotted later in Fig. 5. Note that in Ref. [3] the measured $\chi(T,p)$ data were offset by $+ 0.4$ $10^{-4}$ emu/mole corresponding to the zero of the spin susceptibility determined by NMR. In the present paper we have analyzed the measured data for all samples without making any offsets.

From our analysis of the $\chi(T)$ data so far, it appears that both the $C_B/(T-\theta)$ and the $C/T$ terms are $p$-independent, the latter will be justified again in the following section. Therefore the difference in $\chi(T)T$ for two samples with the same Ca content, but different values of $p$, should arise from changes in their spin susceptibility, $\chi_0$ in Eqn. (2) or equivalently $\chi_{spin}$ in Eqn. (1). This difference, $\Delta\chi T$, will show how the number of QP excitations inside an energy window of width $\sim \varepsilon_F \pm 2k_BT$ changes with $p$. We have gathered other evidence from our earlier charge transport studies on $Y_{1-x}Ca_xBa_2Cu_3O_{7-\delta}$ [4, 5, 10, 13, 25] that a PG does not exist for $p > 0.19 \pm 0.005$. It is therefore convenient to use $\chi(T)T$ for $p > 0.20$ as references. One major advantage of using $\Delta\chi T(p)$ is that we do not need to deal with other contributions to $\chi(T)$. Working with the differences rather than the absolute values practically eliminates the terms that are not related to the quasi-particle DOS.

$\Delta\chi T(p)$ values for the $Y_{1-x}Ca_xBa_2Cu_3O_{7-\delta}$ compounds are shown in Fig. 4 in the temperature range from 100 K to 330 K. Again the lower temperature limit is set to omit the region where significant superconducting fluctuations are present. Fig. 4 highlights a number of interesting and important features, namely, (i) the sudden appearance of the pseudogap at $p \sim 0.19$ and its growth with further underdoping manifested by the decrease in $\Delta\chi T(p)$ with decreasing $p$. This clearly illustrates the loss of QP states near the Fermi level. (ii) $\Delta\chi T(p)$ remains almost unchanged for $p > 0.19$, confirming the absence of a pseudogap for these hole contents. (iii) for values of $p$ that are not too low, a non-zero value of $E_g$ simply gives a constant downward shift in $\Delta\chi T(T,p)$. This is consistent with Eqns. (1) and (3) which also give a constant shift in $\chi T$ (and in $S$) for finite $E_g$ provided $T > 0.25 E_g$. The link between $\chi T$ and $S$ shows that this shift actually arises from the "non-states-conserving" property of the V-shaped DOS. The $\chi T$ curves at lower values of $p$ have a negative slope that is mainly caused by the fact that $T \sim 0.25 E_g$ but other factors such as states in the PG induced by disorder could also be giving a contribution.

Fig. 4a shows $\Delta\chi T(p)$ for pure Y123. Here the hole content of the reference compound is 0.176 and a small PG is present at this composition (see Fig. 5). Nevertheless, $\Delta\chi T(p)$ for pure Y123 shows almost identical behavior to that shown by the Ca substituted samples. This lends further support to our finding that the $C/T$ term from Ca and the $C_B/(T-\theta)$ term from $BaCuO_{2+z}$ impurities are $p$-independent. We have also fitted the $\Delta\chi T(p)$ data for the Ca substituted samples to a V-shaped PG in the DOS, using Eqns. (1) and (3). The $E_g(p)$ values obtained from these fits are also shown in Fig. 5. As found previously [5, 10, 13, 24, 25], $E_g(p)$ is insensitive to the Ca content within the experimental error bars. It can be seen that $E_g(p)$ falls almost linearly with increasing hole concentration, becomes less than $T_c(p)$ on the lightly overdoped side and vanishes for $p > 0.19$.

**Heat Capacity data**

Heat capacity ($C_v$) data for an OD sample from a third-batch of 20%Ca-Y123, with somewhat larger $\chi(T)$ than the OD 20%Ca-Y123 sample from the second batch, are shown in Figs. 6 and 7. They were obtained using a differential technique against a pure copper reference sample. Fig. 6 shows the electronic specific heat coefficient $\gamma(H, T) = C_v/T$ of the sample in magnetic fields from 0 – 13 Tesla after correcting for the difference in phonon terms of the sample and the reference. The influence of magnetic fields ($H$) up to 13 Tesla is demonstrated in Fig. 7 by showing difference plots of $\Delta\gamma(H)$ [$\equiv \gamma(H, T) - \gamma(H = 0, T)$] versus $T$. Over most of the temperature region these difference curves can be scaled to lie on the same curve by dividing them by $Hln(90/H)$, as shown in the inset to Fig. 7. This behavior is predicted by calculations based on the theory of Bardeen, Cooper and Schrieffer for a d-wave superconductor in the dirty limit [26, 27]. It is also given by standard analysis of the vortex lattice in the London limit at intermediate fields, being related by thermodynamics to the $-lnH$ term in the reversible magnetization [28]. However it is surprising that the $HlnH$ scaling still appears to hold above $T_c$, in the range 50 – 65 K. Because the scaled curves can be extrapolated to zero according to an $A-BT^2$ law and have an entropy-conserving property between 70 K (well above $T_c$) and $T \rightarrow 0$, we can use them to determine the contribution, $\gamma_{sc}(H)$, from superconductivity. After subtracting $\gamma_{sc}(H, T)$ from the total electronic term $\gamma(H, T)$ we obtain the residual electronic and magnetic contributions $\gamma_{res}(H, T)$ not associated with the superconducting condensate, that are shown in the inset to Fig. 6. The 13 Tesla data for $\gamma_{res}(H, T)$ extrapolate to a rather low value at $T = 0$ K. This places an upper limit to any electronic contribution due to pair breaking of $\gamma(0) \sim 0.3$ mJ/gat.-K$^2$ at $T = 0$ K. The major part of $\gamma_{res}(H, T)$ below 15 K is therefore of magnetic origin.

Green phase (Y$_2$BaCuO$_5$) impurity shows a sharp specific heat anomaly at ~19 K due to the antiferromagnetic transition [29]. We see no evidence for such an anomaly in the present data, placing an upper limit of less than 1% for this impurity phase. Well oxygenated $BaCuO_{2+z}$ exhibits a very large and strongly field dependent specific heat [23]. Comparison with data for pure $BaCuO_{2+z}$ [23, 30] suggests that the upturn in $\chi(T)$ in zero magnetic field, and approximately half of the weak field dependence, could be explained by 0.038 mole $BaCuO_{2+z}$ impurity per mole Y123, i.e. ~ 1.3% of all Cu atoms being in the form of $BaCuO_{2+z}$. However this could not explain the large Curie term in $\chi(T)$ which would require ~ 0.13 mole of $BaCuO_{2+z}$ or $Y_2BaCuO_5$ per mole Y123, nor the rather large temperature-independent, zero-field contribution to $\gamma_{res}$. We therefore conclude that $\gamma_{res}(H, T)$ results mainly from a magnetic contribution from the Ca ions, supporting our findings from the analysis of $\chi(T)$ data. The weak field dependence that we observe suggests a rather high energy scale of ~ 60 K for these excitations which may reflect a Kondo temperature. Inclusion of such a Kondo temperature in the following high $T$ analysis, i.e. replacing the $C/T$ term in Eqn. (2) by $C/(T+60)$ would increase the values of $C$ obtained from the fits by ca. 30%. However it would not change our conclusions qualitatively and it would make the analysis for $p < 0.19$ less transparent, because as shown in Fig. 4, for $p < 0.19$, the PG gives a negative shift in $\chi(T)T$, equivalent to a negative offset in $C$.

## IV. DISCUSSION

We have found clear evidence that magnetic moments are induced by Ca substitution in Y123 since the increase in $\chi(T)T$ of $Y_{1-x}Ca_xBa_2Cu_3O_{7-\delta}$ with $x$ in Figs. 1 and 2 is too large to be caused by magnetic impurities. Typical fits of our $\chi(T)T$ data to Eqn. (2), multiplied by $T$, for $p = 0.180$ and $p = 0.145$ are shown in Fig. 8, where the values of $C_B$ and $\theta$ for a given Ca content were constrained to be the same as those given in Table 1 for the most overdoped samples. The values of $C$ obtained from these fits are plotted as $C'(x) \equiv C(x) - C(x=0)$, in Fig. 9. We plot the difference $C'(x)$ because $E_g$ is finite for these two values of $p$ and the condition $T > E_g/4$ holds. As discussed already, $E_g$ then gives rise to a $T$-independent negative shift in $\chi(T)T$. If we apply the usual formula $C' = x\, N_{av}\, p_{eff}^2 \mu_B^2 / 3k_B$, where $N_{av}$ is Avogadro's number and $\mu_B$ the Bohr magneton, then the effective moment per $Ca^{++}$ ion, $p_{eff} = 1.35 \pm 0.05$ (in the units of $\mu_B$) irrespective of $p$. The origin of this Ca induced magnetic moment is not entirely clear. As stated earlier, we think it is unlikely that this is related to the proposed appearance of in-plane oxygen vacancies with magnetic character [19]. In-plane disorder capable of giving such a large Curie term in the magnetic susceptibility should reduce $T_{cmax}$ more drastically. For example 20%Ca substitution only reduces the maximum $T_c$, $T_{cmax}$, of pure Y123 by ~ 9 K, equivalent to ~ 1% Zn substitution in the $CuO_2$ plane.

As shown in Fig. 9, $C'(x)$ is only approximately linear. Moreover, comparison of the $C_{ESR}$ and $C_B$ values in Table 1 suggests that we may have underestimated $C_B$ for $x = 0.05$ and 0.1 causing the corresponding $C'$ values in Fig. 9 to be $3 \times 10^{-3}$ emu-K/mole too high. So the true behavior of $C'(x)$ could be flatter at low $x$. Therefore we examine two-possible scenarios: *(a)* All $Ca^{2+}$ give a smaller effect with $p_{eff}$ ~ 1.35 (less than $p_{eff}$ ~ 1.7 for spin $s = \frac{1}{2}$). In this case $C'(x)$ would be linear and furthermore it is hard to reconcile the presence of localized spins induced by Ca, presumably in the $CuO_2$ planes, with the mobile carriers that Ca definitely induces in the $CuO_2$ planes [11, 31]. This puzzle can be resolved by scenario *(b)* An isolated $Ca^{2+}$ does not localize carriers in its vicinity, gives no magnetic moment and donates one extra mobile carrier in the $CuO_2$ planes, but statistical clusters of two or more nearest-neighbour $Ca^{2+}$ ions cause a stronger perturbation in the neighboring $CuO_2$ planes and give rise to a magnetic moment. For $x = 0.20$, the concentration of isolated $Ca^{2+}$ ion is given by $x(1-x)^4 = 0.082$ and that for non-isolated $Ca^{2+}$ is $x[1-(1-x)^4] = 0.118$. In this model the data in Fig. 9 give $p_{eff}$ ~ 1.7, the value expected for $s = \frac{1}{2}$. Scenario *(b)* is further supported by the following points. (i) As Fig. 9 shows, the Ca induced Curie constant and the probability of finding non-isolated $Ca^{2+}$ have similar $x$-dependences. (ii) As shown in Fig. 10, $T_{cmax}$ vs. $x$ has a downward curvature possibly because of increased scattering from the stronger perturbations in the $CuO_2$ planes. (iii) This 'pair' picture is consistent with experimental evidence that Ca is less effective as a hole donor for larger $x$. For example, for nearly fully oxygenated compounds, the maximum planar hole contents are 0.203, 0.217, and 0.236 ($\pm$ 0.004) for $x = 0.05$, 0.1 and 0.2, respectively [10]. Whereas from simple electron counting $Ca^{2+}$ is expected to add $x/2$ holes per $CuO_2$ plane, giving 0.205, 0.23, and 0.28 for the corresponding $x$ values. It should be mentioned that a similar model was proposed by Hammel *et al.* [32] on the basis of their NMR measurements, when considering possible hole localization in $La_2CuO_{4+y}$ and $La_{2-x}Sr_xCuO_4$.

Analysis of $\Delta\chi T(p)$ data complements our earlier analysis of resistivity, $\rho(T, p)$, data quite well [13, 25, 33, 34]. We see quite clearly from $\Delta\chi T(p)$ that the PG vanishes abruptly at $p \sim 0.19$.

## V. CONCLUSIONS

In conclusion, we have reported a systematic study of the static magnetic susceptibility for polycrystalline $Y_{1-x}Ca_xBa_2Cu_3O_{7-\delta}$. We have found a Curie-like contribution due to Ca substitution that is almost independent of $p$ and oxygen content and put forward a model where this arises from statistical clusters of two or more nearest neighbor Ca atoms. We have also seen from the analysis of $\chi(T)$ data that the pseudogap vanishes abruptly for $p > 0.19$. As $p$ is reduced below 0.19, $\chi(T)T$ decreases, and therefore the PG energy $E_g$ must increase. Within models involving preformed pairs [2], and others in which the PG sets in abruptly below a certain temperature one would expect $\chi(T)T$ vs. $T$ plots for samples with a PG to merge with each other at higher temperatures where the pseudogap is zero. We have not seen any sign of a recovery in $\chi(T)T$ up to 400 K (the upper temperature limit of the measurements). This implies that the pseudogap causes a permanent loss of states near the Fermi level and that it does not close as the temperature rises. The same conclusions were reached from earlier specific heat measurements [3 - 5]. This places severe constraints on possible theories of the PG.

## ACKNOWLEDGEMENTS

We would like to thank Prof. J. L. Tallon for preparing some of the samples and for helpful suggestions and comments on many occasions over a long period. We also thank Prof. P. Monod, Dr. K. Yates, Dr. C. Hayward, Ms. Y. Shi, and Mr. D. M. Astill for making the ESR, Raman and EPMA measurements. One of us (SHN) acknowledges financial support from the Commonwealth Scholarship Commission (UK), Darwin College, the Cambridge Philosophical Society, the Lundgren Fund, the Department of Physics and the IRC in Superconductivity, Cambridge. SHN also thanks the Quantum Matter group, University of Cambridge for their hospitality.

*Corresponding author. E-mail: salehnaqib@yahoo.com

Table 1: Sample parameters including those from fits to $\chi(T)T = \chi_0 T + C + C_B T/(T-\theta)$ from 100 – 330 K. The $\equiv$ sign means the parameter was fixed at the value shown, n/a means not applicable. Also shown are the contributions ($C_{ESR}$) to the Curie constant for samples with $p = 0.16$, arising from unwanted paramagnetic phases detected by ESR. The * symbol shows samples for which there was a ferromagnetic background at the time of the ESR measurements, giving larger uncertainty in $C_{ESR}$.

| Sample Ca % | EPMA $x$(Ca) | $p$ holes/Cu | $\chi_0$ $10^{-4}$ emu/mole | $C$ $10^{-4}$ emu-K/mole | $C_B$ $10^{-4}$ emu-K/mole | $\theta$ K | $C_{ESR}$ $10^{-4}$ emu-K/mole |
|---|---|---|---|---|---|---|---|
| 0 | - | 0.176 | 2.71 | -11 | $\equiv 0$ | n/a | n/a |
| 5 | 0.049±0.006 | 0.203 | 2.92 | 49 | $\equiv 0$ | n/a | 34* |
| 10 | 0.105±0.010 | 0.220 | 2.42 | 263 | 6.8 | $\equiv 62$ | 34±4 |
| 20 I | 0.195±0.012 | 0.220 | 2.75 | 398 | 25 | $\equiv 62$ | 30* |
| 20 II | 0.20±0.03 | 0.215 | 2.30 | 492 | 68 | $\equiv 62$ | 82±11 |
| 20 III | - | 0.218 | 2.56 | 481 | $\equiv 146$ | $\equiv 62$ | n/a |

**Figure Captions:**

Figure 1: $\chi(T)$ and $\chi(T)T$ for (a) YBa$_2$Cu$_3$O$_{7-\delta}$, and (b) Y$_{0.80}$Ca$_{0.20}$Ba$_2$Cu$_3$O$_{7-\delta}$ (first batch). *p*-values are shown and are accurate to ± 0.004.

Figure 2: $\chi(T)$ and $\chi(T)T$ for (a) Y$_{0.95}$Ca$_{0.05}$Ba$_2$Cu$_3$O$_{7-\delta}$, (b) Y$_{0.90}$Ca$_{0.10}$Ba$_2$Cu$_3$O$_{7-\delta}$, and (c) Y$_{0.80}$Ca$_{0.20}$Ba$_2$Cu$_3$O$_{7-\delta}$ (second batch). *p*-values are shown and are accurate to ± 0.004.

Figure 3: Fits (full lines) of normal-state $\chi(T, p)$ data [3] for polycrystalline YBa$_2$Cu$_3$O$_{7-\delta}$, with the oxygen deficiencies $\delta$ shown, to Eqns. 1 and 3 in text. The parameter $N_0$ was fixed at a value corresponding to $\chi = 2.75 \; 10^{-4}$ emu/mole for $E_g = 0$. Values of $E_g$ obtained from the fits are given in the figure.

Figure 4 (Color online): $\Delta\chi T(p)$ $[\equiv \chi T(p) - \chi T(p_{ref})]$ for (a) pure Y123, $p_{ref} = 0.176$. (b) Y$_{0.95}$Ca$_{0.05}$Ba$_2$Cu$_3$O$_{7-\delta}$, $p_{ref} = 0.203$, (c) Y$_{0.90}$Ca$_{0.10}$Ba$_2$Cu$_3$O$_{7-\delta}$. $p_{ref} = 0.201$, (d) Y$_{0.80}$Ca$_{0.20}$Ba$_2$Cu$_3$O$_{7-\delta}$, $p_{ref} = 0.215$ (second-batch) with one data set at $p = 0.113$ for the first batch with $p_{ref} = 0.220$. $p$-values are accurate to $\pm 0.004$.

Figure 5 (Color online): $E_g(p)$ for Y$_{1-x}$Ca$_x$Ba$_2$Cu$_3$O$_{7-\delta}$ compounds. The parabolic $T_c(p)$ curve with $T_{cmax} = 93$ K is also shown. The dashed-line is a guide to the eye.

Figure 6: Main - magnetic field and temperature dependence of the electronic specific heat coefficient, $\gamma(H, T)$ for a third-batch of OD Y$_{0.80}$Ca$_{0.20}$Ba$_2$Cu$_3$O$_{7-\delta}$. Inset - the residual specific heat coefficient, $\gamma_{res}(H, T)$ after subtracting the contribution from superconductivity.

Figure 7 (Color online): Main - difference in specific heat coefficients $[\gamma(H, T) - \gamma(H = 0, T)]$ versus $T$ for the third-batch of OD Y$_{0.80}$Ca$_{0.20}$Ba$_2$Cu$_3$O$_{7-\delta}$. Inset - scaling of the difference plot used to determine the contribution from superconductivity. The thick line represents the data at 13 Tesla.

Figure 8 (Color online): (a) $\chi(T)T$ for Y$_{1-x}$Ca$_x$Ba$_2$Cu$_3$O$_{7-\delta}$ with $p = 0.180 \pm 0.005$ and (b) Y$_{1-x}$Ca$_x$Ba$_2$Cu$_3$O$_{7-\delta}$ with $p = 0.145 \pm 0.005$. Ca contents ($x$) are shown in the figure. Full (red) lines are fits to Eqn. 2 (multiplied by $T$) with $C_B$ and $\theta$ values given in Table 1. The values of $C$ obtained from these fits are plotted as $C'(x) \equiv C(x) - C(x=0)$ vs. $x$ in Fig. 9.

Figure 9 (Color online): $C'$ and the probability of non-isolated Ca atoms (dashed line) versus Ca content ($x$) determined by EPMA, for Y$_{1-x}$Ca$_x$Ba$_2$Cu$_3$O$_{7-\delta}$ with $p = 0.180$ and 0.145. $C'$ is the difference, $C(x) - C(x=0)$, at the same value of $p$.

Figure 10 (Color online): The maximum $T_c$ ($\equiv T_{cmax}$) versus Ca content ($x$) in Y$_{1-x}$Ca$_x$Ba$_2$Cu$_3$O$_{7-\delta}$. The dashed line is a guide to the eye.

Figure 1:

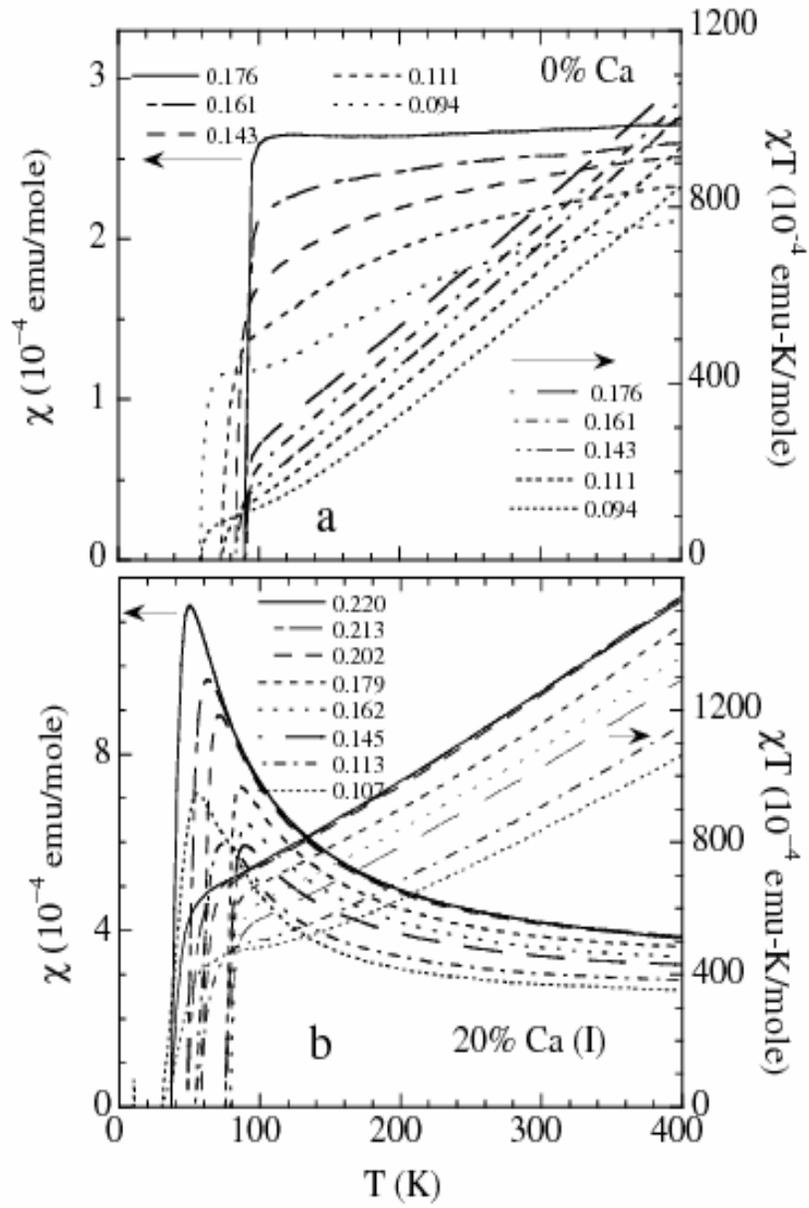

Figure 2:

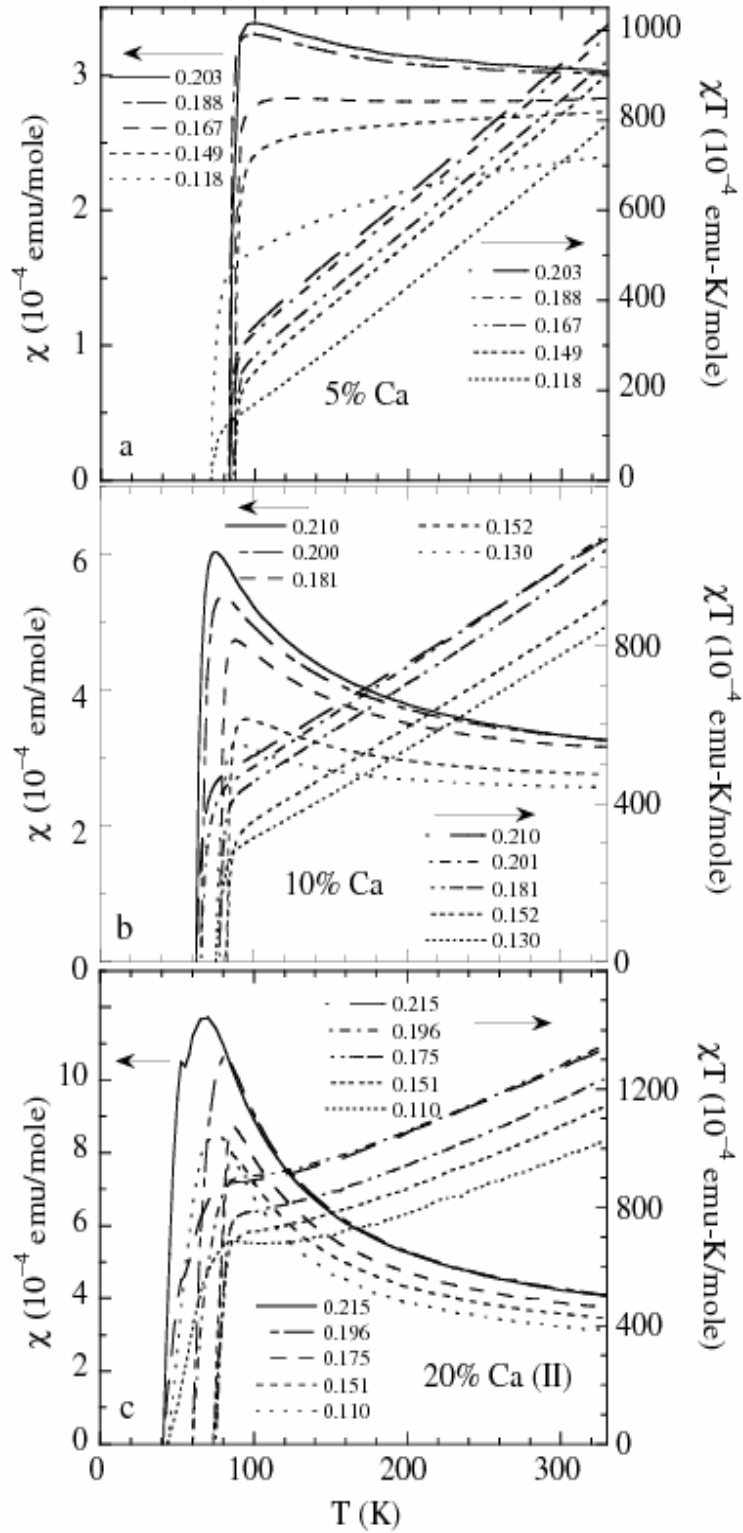

Figure 3:

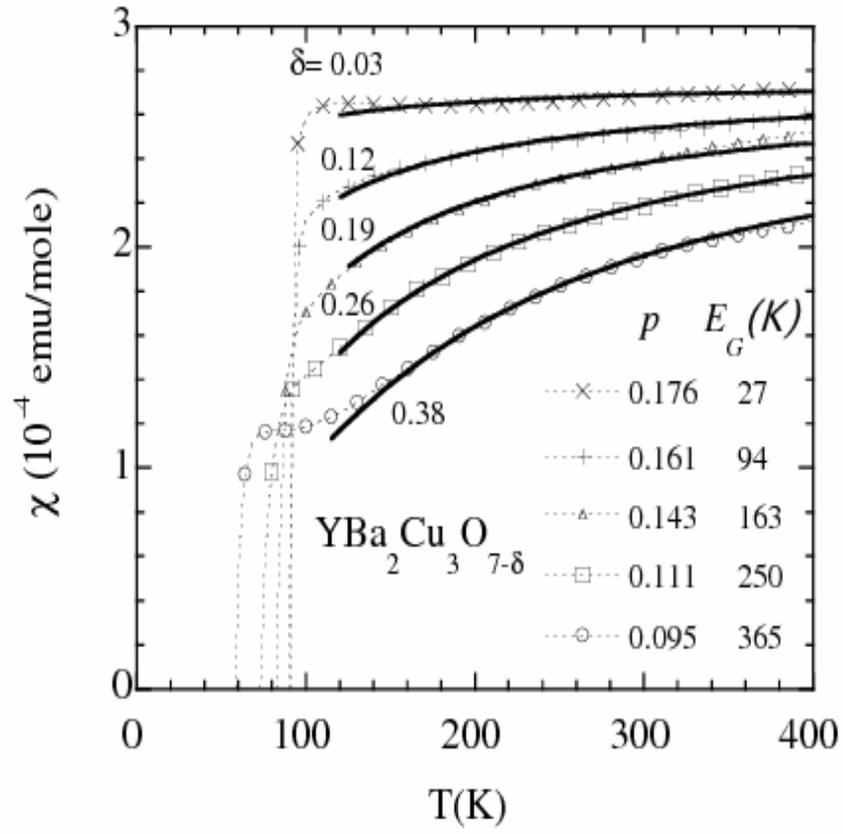

Figure 4:

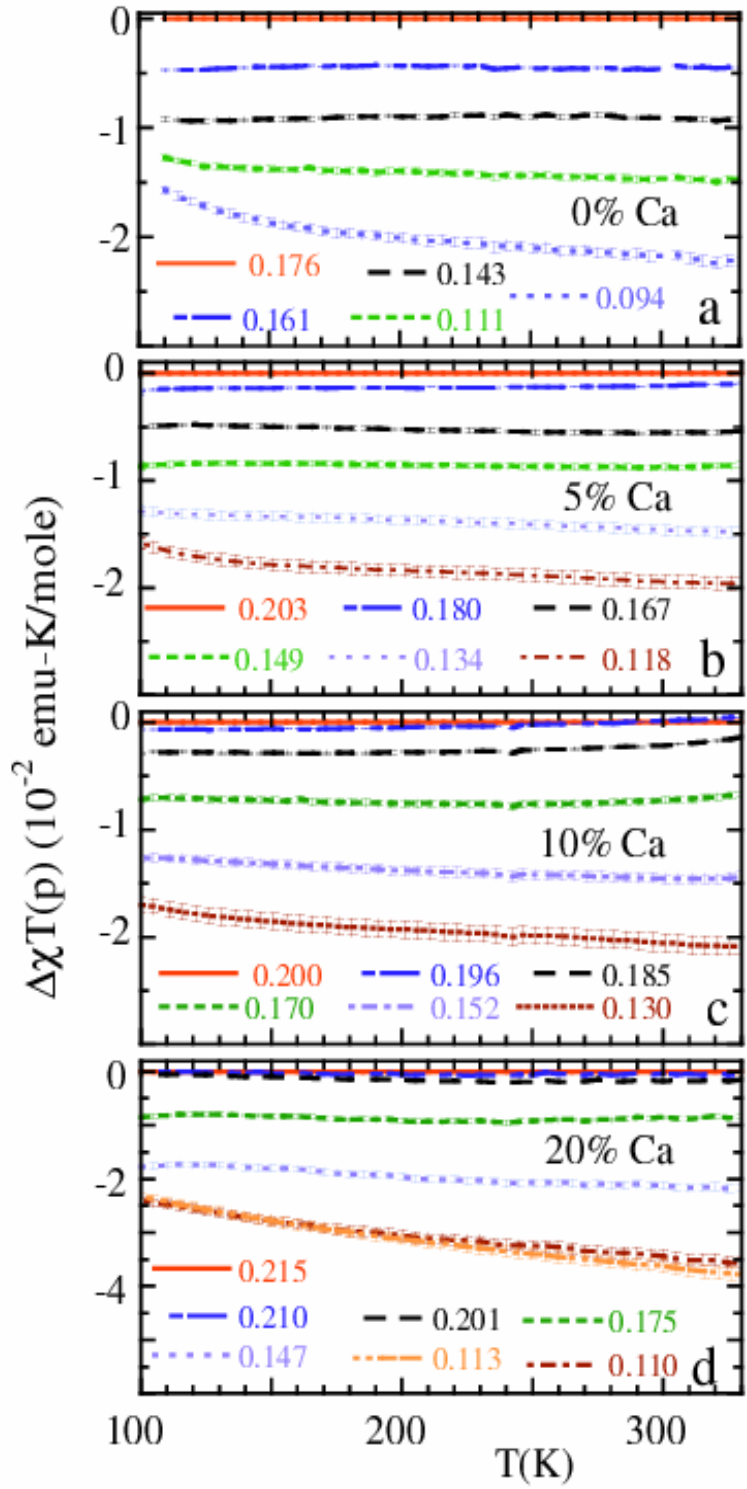

Figure 5:

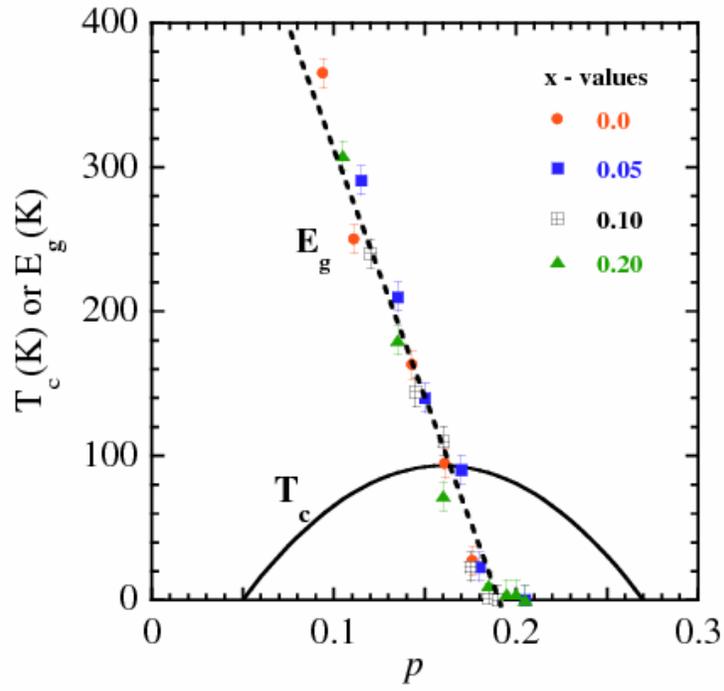

Figure 6:

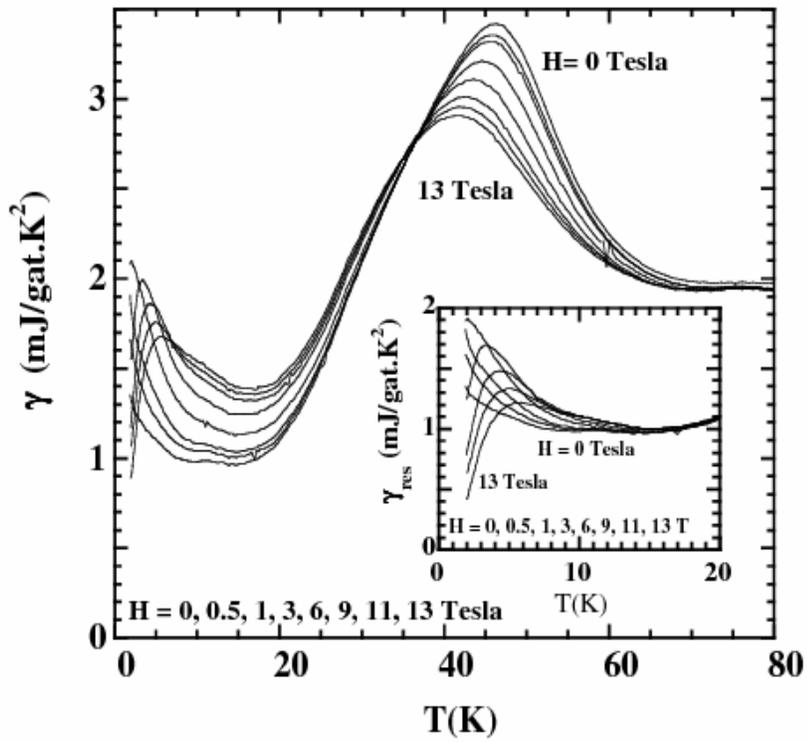

Figure 7:

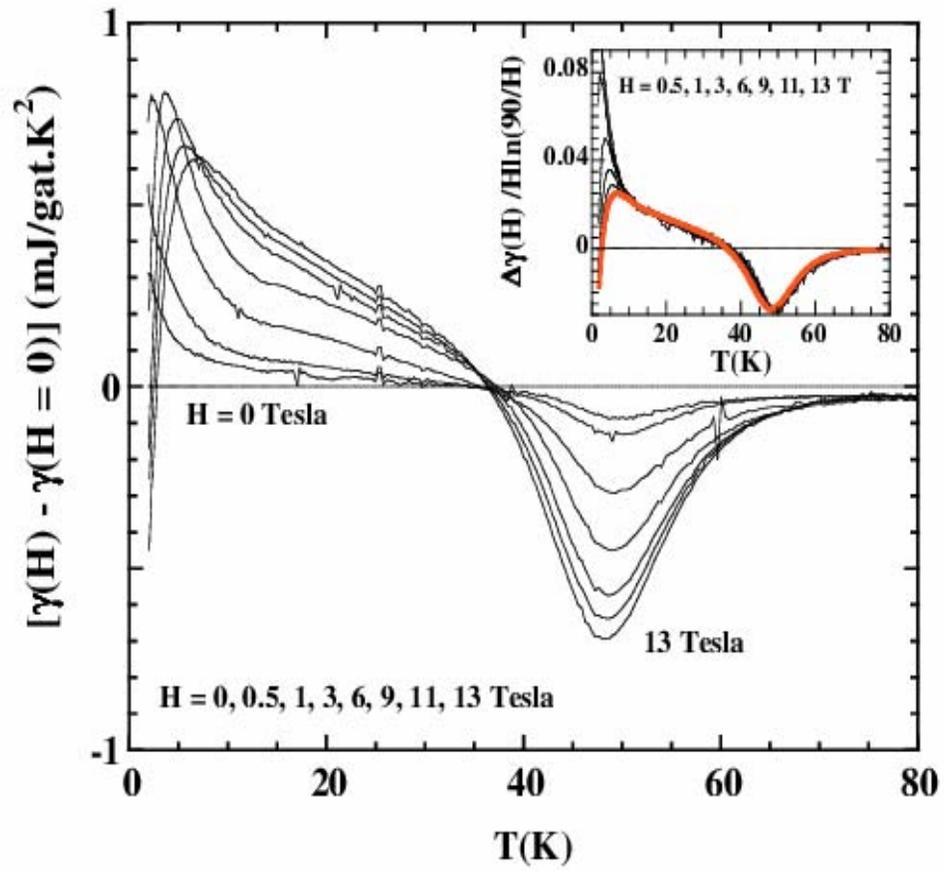

Figure 8:

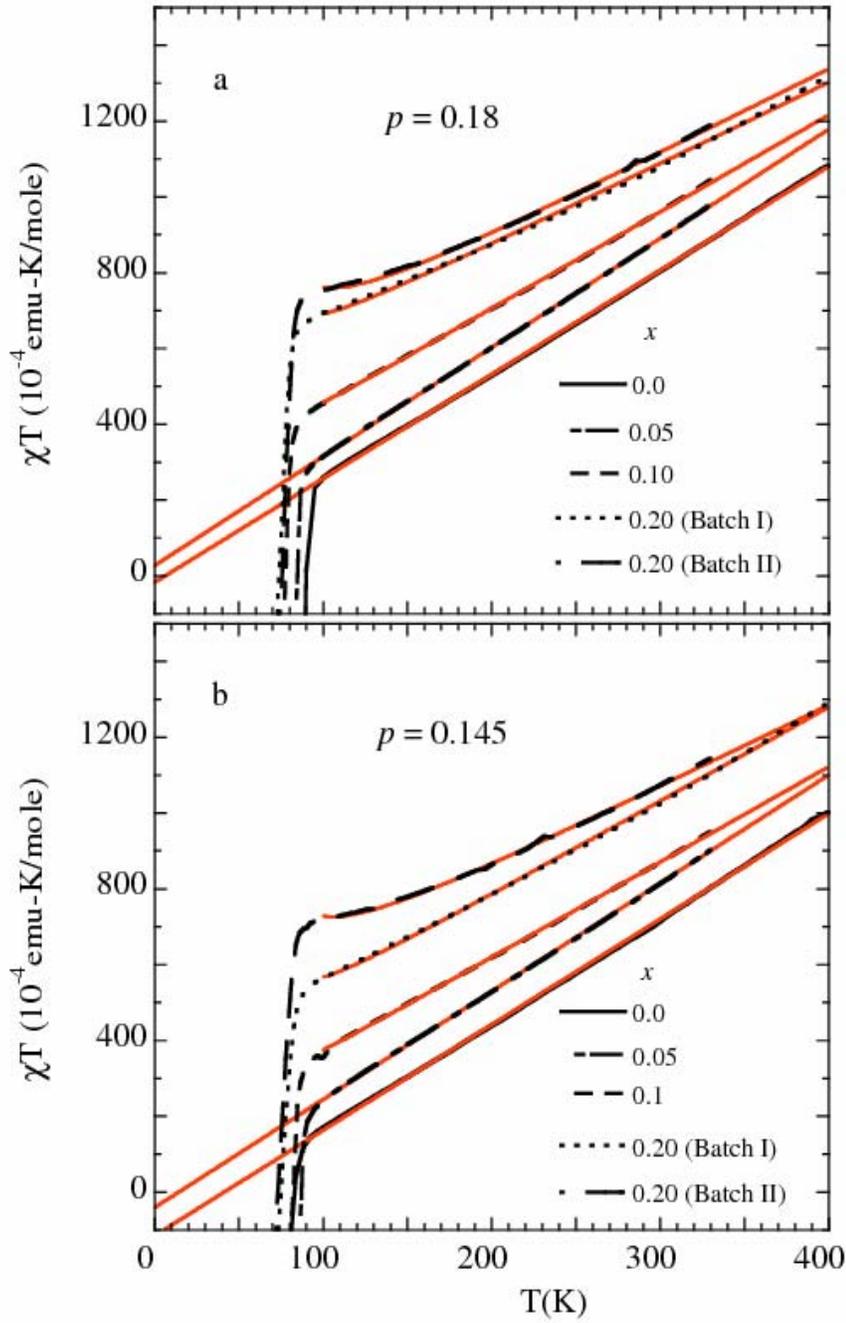

Figure 9:

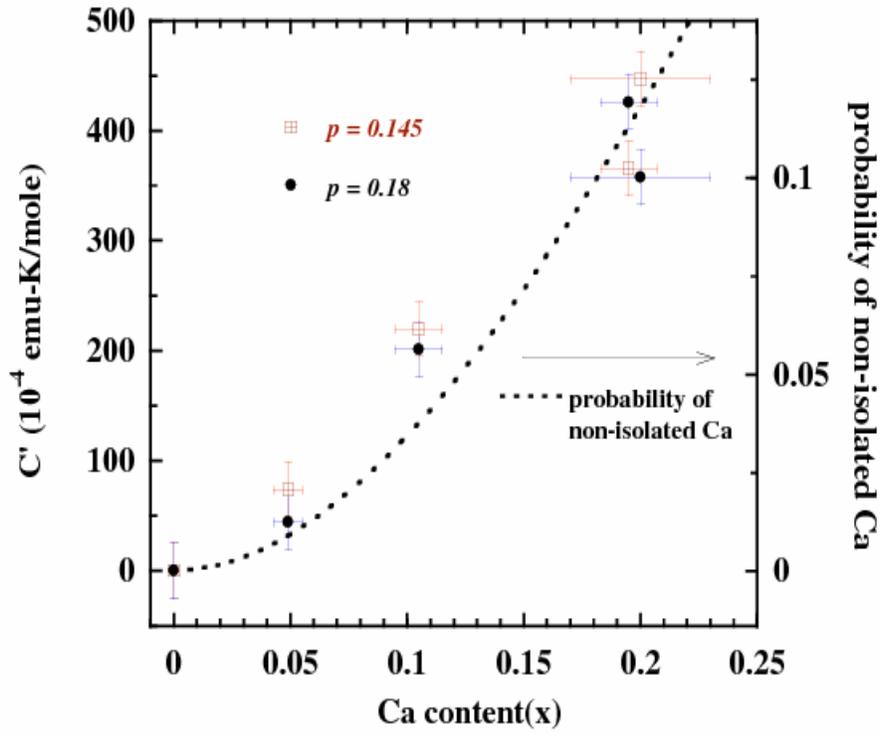

Figure 10:

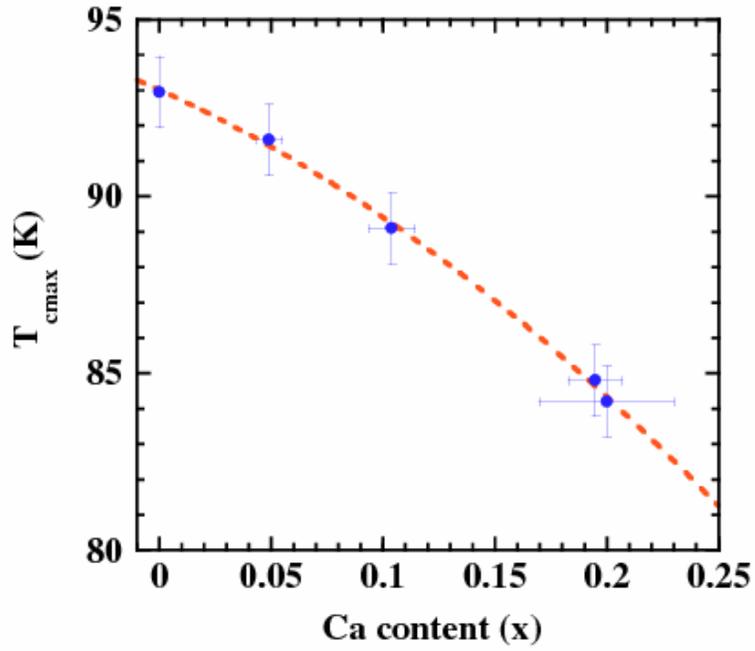